\documentclass[12pt]{article}

\usepackage{times}
\usepackage[german,english]{babel}
\usepackage{dbairep}

\foreignreport

\authornames{
Frank Harary, Wolfgang Slany, and Oleg Verbitsky
}

\published{A preliminary version of this paper appeared in 
the
          Proceedings of MSRI-CGTR'2000, the 2nd Combinatorial Game Theory
          Research Workshop at the Mathematical Sciences Research Institute in
          Berkeley, California.\\
Online version of this paper:
http://www.dbai.tuwien.ac.at/staff/slany/pubs/dbai-tr-2001-42.ps.gz}

\abstract{
\noindent In the graph avoidance game two players alternatingly color edges of a
graph $G$ in red and in blue respectively. The player who first
creates a monochromatic subgraph isomorphic to a forbidden graph $F$
loses.  A \emph{symmetric strategy} of the second player ensures that,
independently of the first player's strategy, the blue and the red
subgraph are isomorphic after every round of the game.  We address
the class of those graphs $G$ that admit a symmetric strategy for all
$F$ and discuss relevant graph-theoretic and complexity issues.  We
also show examples when, though a symmetric strategy on $G$ generally
does not exist, it is still available for a particular $F$.}

\infsysrr{DBAI-TR-2001-42}
\date{September 30, 2001}

\usepackage{epic}
\usepackage{eepic}

\newcommand{\hide}[1]{}

\def\emline#1#2#3#4#5#6{%
      \path(#1,#2)(#4,#5)}

\usepackage{latexsym}
\newtheorem{thm}{Theorem}[section]
\newenvironment{theorem}{\begin{thm}\begin{slshape}}{\end{slshape}\end{thm}}
\newtheorem{dfn}[thm]{Definition}
\newenvironment{definition}{\begin{dfn}\em}{\end{dfn}}
\newtheorem{rem}[thm]{Remark}
\newenvironment{remark}{\begin{rem}\em}{\end{rem}}
\newtheorem{que}[thm]{Question}
\newenvironment{question}{\begin{que}\em}{\end{que}}
\newtheorem{xmp}[thm]{Example}
\newenvironment{example}{\begin{xmp}\em}{\end{xmp}}

\newcommand{\qed}{$\;\;\;\Box$}
\newenvironment{proof}{\par\smallbreak{\noindent\sl Proof.~}}
{\unskip\nobreak\hfill \qed \par\medbreak}

\newcommand{\one}{$\cal{A}$}
\newcommand{\two}{$\cal{B}$}

\newcommand{\V}{V}
\newcommand{\E}{E}
\newcommand{\avoid}[2]{\mbox{AVOID}( #1, #2)}
\newcommand{\pr}{\mathbin{{\otimes}_1}}
\newcommand{\prr}{\mathbin{{\otimes}_2}}
\newcommand{\prrr}{\mathbin{{\otimes}_3}}
\newcommand{\pri}{\mathbin{{\otimes}_i}}
\newsavebox{\bsymm}
\newsavebox{\bwin}
\newsavebox{\bauto}
\newsavebox{\blauto}
\savebox{\bsymm}{$\mathcal{C}_{\mathrm{sym}}$}
\savebox{\bwin}{$\mathcal{C}_{\mathrm{II}}$}
\savebox{\bauto}{$\mathcal{C}_{\mathrm{auto}}$}
\savebox{\blauto}{$\mathcal{C}_{\mathrm{line-auto}}$}
\newcommand{\symm}{\usebox{\bsymm}}
\newcommand{\win}{\usebox{\bwin}}
\newcommand{\auto}{\usebox{\bauto}}

\newcommand{\Lubiw}{{\sc Order 2 Fixed-Point-Free Automorphism}}
\newcommand{\PAR}{{\sc PAR}}
\newcommand{\GI}{{\sc GI}}

\newcommand{\function}[2]{:#1 \rightarrow #2}

\title{A Symmetric Strategy\\
in Graph Avoidance Games}

\author{Frank~Harary\affiliation{Computer Science Department,
New Mexico State University,
Las Cruces, NM 88003, USA.}\and
Wolfgang~Slany\affiliation{Institut f\"ur Informationssysteme, Technische Universit\"at Wien,
Favoritenstra\ss{}e 9, A-1040 Wien, Austria.}\thanks{Research partly supported by
Austrian Science Foundation grant Z29-INF.}\and
Oleg~Verbitsky\affiliation{Department of Mechanics \& Mathematics, Lviv University,
Universytetska~1, 79000 Lviv, Ukraine.}\thanks{Research was partly done while visiting the
Institut f\"ur Informationssysteme at the Technische Universit\"at Wien,
supported by a Lise Meitner Fellowship of the Austrian
Science Foundation (FWF grant M~532).}}

\begin{document}
\maketitle

\hide{
\begin{abstract}
In the graph avoidance game two players alternatingly color edges of a
graph $G$ in red and in blue respectively. The player who first
creates a monochromatic subgraph isomorphic to a forbidden graph $F$
loses.  A \emph{symmetric strategy} of the second player ensures that,
independently of the first player's strategy, the blue and the red
subgraph are isomorphic after every round of the game.  We address
the class of those graphs $G$ that admit a symmetric strategy for all
$F$ and discuss relevant graph-theoretic and complexity issues.  We
also show examples when, though a symmetric strategy on $G$ generally
does not exist, it is still available for a particular $F$.
\end{abstract}
}

\section{Introduction}

In a broad class of games that have been studied in the literature,
two players, \one\/ and \two, alternately color edges of a graph $G$
in red and in blue respectively. In the achievement game the objective is to
create a monochromatic subgraph isomorphic to a given graph $F$. In
the avoidance game the objective is, on the contrary, to avoid
creating such a subgraph.  Both the achievement and the avoidance
games have strong and weak versions.  In the strong version \one\/ and
\two\/ both have the same objective.  In the weak version \two\/ just
plays against \one, that is, tries either to prevent \one\/ from
creating a copy of $F$ in the achievement game or to force such
creation in the avoidance game.  The weak achievement game, known also
as the Maker-Breaker game, is most studied \cite{ESe,Bec,Pek}. Our
paper is motivated by the strong avoidance game \cite{Har,EHa} where
monochromatic $F$-subgraphs of $G$ are forbidden, and the player who
first creates such a subgraph loses.

The instance of a strong avoidance game with $G=K_6$ and
$F=K_3$ is well known under the name SIM \cite{Sim}.
Since for any bicoloring of $K_6$ there is a monochromatic $K_3$,
a draw in this case is impossible. It is proven in \cite{MRH}
that a winning strategy in SIM is available for \two.
A few other results for small graphs are known \cite{Har}.
Note that, in contrast with the weak achievement games,
if \two\/ has a winning strategy in the avoidance game on $G$
with forbidden $F$ and if $G$ is a subgraph of $G'$, then it is not
necessary that \two\/ also has a winning strategy on $G'$ with forbidden $F$.
%Example: F=K_{1,2}, G=K_3+e, G'=G+e at the vertex of degree 3
Recognition of a winner seems generally to be
a non-trivial task both from the combinatorial and from the
complexity-theoretic point of view (for complexity issues
see, e.g., \cite{Slany99}).

In this paper we introduce the notion of a {\em symmetric
strategy\/}\footnote{Note that this term has been
used also in other game-theoretic situations (see, e.g., \cite{RGo}).}
for \two. We say that \two\/ follows a symmetric strategy on $G$ if
after every move of \two\/ the blue and the red subgraphs are isomorphic,
irrespective of \one's strategy. As easily seen, if \two\/
plays so, he at least does not lose in
the avoidance game on $G$ with any forbidden $F$. There is a similarity
with the {\em mirror-image strategy\/} of \one\/ in the achievement
game~\cite{BCG}. However, the latter strategy is used on two disjoint
copies of the complete graph, and therefore in our case things are much
more complicated.

We address the class $\symm$ of those graphs $G$ on which
a symmetric strategy for \two\/ exists. We observe that
$\symm$ contains all graphs having an involutory automorphism
without fixed edges. This subclass of $\symm$, denoted by $\auto$,
includes even paths and cycles, bipartite complete graphs $K_{s,t}$
with $s$ or $t$ even, cubes, and the Platonic graphs except the tetrahedron.
We therefore obtain a lot of instances of the avoidance game with a winning
strategy for \two. More instances can be obtained based on closure
properties of $\auto$ that we check with respect to a few basic graph
operations.

Nevertheless, recognizing a suitable automorphism and, therefore,
using the corresponding symmetric strategy is not easy.
Based on a related result of Lubiw \cite{Lub}, we show that
deciding membership in $\auto$ is NP-complete.

We then focus on games on complete graphs.  We show that $K_n$ is not
in $\symm$ for all $n\ge 4$.  Moreover, for an arbitrary strategy of
\two, \one\/ is able to violate the isomorphism between the red and
the blue subgraphs in at most $n-1$ moves. Nevertheless, we consider
the avoidance game on $K_n$ with forbidden $P_2$, a path of length 2,
and point out a simple symmetric strategy making \two\/ the winner.
This shows an example of a graph $G$ for which, while a
symmetric strategy in the avoidance game does not exist in general, it
does exist for a particular forbidden $F$.

The paper is organized as follows. Section \ref{s:defn} contains the
precise definitions. In Section \ref{s:list} we compile the membership
list for $\symm$ and $\auto$. In Section \ref{s:close} we investigate
the closure properties of $\symm$ and $\auto$ with respect to various
graph products. In Section \ref{s:np} we prove the NP-completeness of
$\auto$. Section \ref{s:p_2} analyses the avoidance game on $K_n$ with
forbidden $P_2$.

\section{Definitions}\label{s:defn}

We deal with two-person positional games of the following kind.
Two players, \one\/ and \two, alternatingly color edges of a graph
$G$ in red and in blue respectively. Player \one\/ starts the game.
In a {\em move\/} a player colors an edge that was so-far uncolored.
The $i$-th {\em round\/} consists of the $i$-th move of \one\/
and the $i$-th move of \two.
Let $a_i$ (resp.\ $b_i$) denote an edge
colored by \one\/ (resp.\ \two) in the $i$-th round.

A {\em strategy\/} for a player determines the edge to be colored at every
round of the game. Formally, let $\epsilon$ denote the empty sequence.
A strategy of \one\/ is a function $S_1$ that maps every possibly empty
sequence of pairwise distinct edges $e_1,\ldots,e_i$
into an edge different from $e_1,\ldots,e_i$ and from
$S_1(\epsilon),S_1(e_1),\allowbreak
S_1(e_1,e_2),\ldots,S_1(e_1,\ldots,e_{i-1})$.
A strategy of \two\/ is a function $S_2$ that maps every nonempty
sequence of pairwise distinct edges $e_1,\ldots,e_i$
into an edge different from $e_1,\ldots,e_i$ and from
$S_2(e_1),S_2(e_1,e_2),\ldots,\allowbreak S_2(e_1,\ldots,e_{i-1})$.
If \one\/ follows a strategy $S_1$ and \two\/ follows a strategy $S_2$,
then $a_i=S_1(b_1,\ldots,b_{i-1})$ and $b_i=S_2(a_1,\ldots,a_{i})$.

Let $A_i=\{a_1,\ldots,a_i\}$ (resp.\ $B_i=\{b_1,\ldots,b_i\}$) consist of
the red (resp.\ blue) edges colored up to the $i$-th round.
A {\em symmetric strategy of \two\/ on $G$\/} ensures that,
irrespective of \one's strategy, the subgraphs $A_i$ and $B_i$ are
isomorphic for every $i\le m/2$, where $m$ is the size of $G$.

The class of all graphs $G$ on which \two\/ has a
symmetric strategy will be denoted by $\symm$.

Suppose that we are given graphs $G$ and $F$ and that $F$ is
a subgraph of $G$. The {\em avoidance game
on $G$ with a forbidden subgraph $F$\/} or, shortly, the game $\avoid G F$
is played as described above with the following ending condition:
The player who first creates a monochromatic subgraph of $G$
isomorphic to $F$ loses.

Observe that a symmetric strategy of \two\/
on $G$ is non-losing for \two\/ in $\avoid G F$, for every
forbidden $F$. Really, the assumption that \two\/ creates a
monochromatic copy of $F$ implies that such a copy is already
created by \one\/ earlier in the same round.

\section{Automorphism-based strategy}\label{s:list}

Given a graph $G$, we denote its vertex set by $\V(G)$ and its edge set
by $\E(G)$.
An {\em automorphism\/} of a graph $G$ is a permutation of $\V(G)$ that
preserves the vertex adjacency. Recall that the {\em order\/} of
a permutation is the minimal $k$ such that the $k$-fold composition
of the permutation is the identity permutation. In particular,
a permutation of order 2, also called an {\em involution\/},
coincides with its inversion.
We call an automorphism of order 2 {\em involutory}.
%We write $G\cong H$ if graphs $G$ and $H$ are isomorphic.

The symmetric strategy can be realized if a graph $G$
has an involutory automorphism that moves every edge.
More precisely, an automorphism $\phi\function{\V(G)}{\V(G)}$
determines a permutation $\phi'\function{\E(G)}{\E(G)}$ by
$\phi'(\{u,v\})=\{\phi(u),\phi(v)\}$. We assume that $\phi$
is involutory and $\phi'$ has no fixed element.
In this case, whenever \one\/ chooses an edge $e$, \two\/ chooses
the edge $\phi'(e)$.
This strategy of \two\/ is well defined because $\E(G)$
is partitioned into 2-subsets of the form $\{e,\phi'(e)\}$. This strategy is
really symmetric because after completion of every round $\phi$ induces
an isomorphism between the red and the blue subgraphs.
We will call such a strategy {\em automorphism-based}.

\begin{definition}
$\auto$ is a subclass of $\symm$ consisting of all those graphs $G$ on
which \two\/ has an automorphism-based symmetric strategy.
\end{definition}

We now list some examples of graphs in $\auto$.
\begin{example}\label{ex:auto}
{\sl Graphs in $\auto$.}
\begin{enumerate}
\item
$P_n$, a path of length $n$, if $n$ is even.
\item
$C_n$, a cycle of length $n$, if $n$ is even.
\item
Four Platonic graphs excluding the tetrahedron.
\item
Cubes of any dimension.\footnote{More generally, cubes are a particular
case of grids, i.e., Cartesian products of paths. The central
symmetry of a grid moves each edge unless exactly one of the factors
is an odd path.}
\item
Antipodal graphs (in the sense of \cite{BKS}) of size more than 1.
Those are connected graphs such that for every vertex $v$, there is a
unique vertex $\bar v$ of maximum distance from $v$. The
correspondence $\phi(v)=\bar v$ is an automorphism \cite{Kot}.  As
easily seen, it is involutory and has no fixed edge. The class of
antipodal graphs includes the graphs from the three preceding items.
\item
$K_{s,t}$, a bipartite graph whose classes have $s$ and $t$ vertices,
if $st$ is even.
\item
$K_{s,t}-e$, that is, $K_{s,t}$ with an edge deleted, provided $st$ is odd.
\item
$K_n$, a complete graph on $n$ vertices, with a matching
of size $\lfloor n/2\rfloor$ deleted.
Note that in this and the preceding examples, for all choices of edges to
be deleted, the result of deletion is the same up to an isomorphism.
\end{enumerate}
\end{example}

%\begin{example}\label{ex:iso}
%The union $G\cup G'$ where $G$ and $G'$ are arbitrary isomorphic graphs
%on disjoint vertex sets.
%Generally, if $k$ is even, then $kG$ is in $\auto$, while if
%$k$ is odd, the outcome of any game $(kG,F)$ is the same as in the game $(G,F)$.
%\end{example}

It turns out that a symmetric strategy is not necessarily
auto\-mor\-phism-based.

\begin{theorem}
$\auto$ is a proper subclass of $\symm$.
\end{theorem}
Below is a list of a few separating examples.
\begin{example}\label{ex:notauto}
{\sl Graphs in $\symm\setminus\auto$.}
\begin{enumerate}
\item
A triangle with one more edge attached
(the first graph in Figure~\ref{fig:symgraphs}).
This is the only connected separating example of even size we know.
In particular, none of the connected graphs of size 6 is in
$\symm\setminus\auto$. Note that the definition of $\symm$ does not
exclude graphs of odd size, as given in the further examples.
\item
The graphs of size 5 shown in Figure~\ref{fig:symgraphs}.
\item
Paths $P_1$, $P_3$, and $P_5$.
\item
Cycles $C_3$, $C_5$, and $C_7$.
\item
Stars $K_{1,n}$, if $n$ is odd.
\end{enumerate}
\end{example}

Note that in spite of items 4 and 5, $P_7$ and $C_9$ are not in $\symm$.

\begin{figure}
\centerline{
\unitlength=1.00mm
%\special{em:linewidth 0.4pt}
\linethickness{0.4pt}
\begin{picture}(112.00,66.00)
\put(5.00,45.00){\circle*{2.00}}
\put(15.00,45.00){\circle*{2.00}}
\put(80.00,30.00){\circle*{2.00}}
\put(75.00,25.00){\circle*{2.00}}
\put(85.00,25.00){\circle*{2.00}}
\put(80.00,20.00){\circle*{2.00}}
\put(80.00,10.00){\circle*{2.00}}
\put(10.00,35.00){\circle*{2.00}}
\put(10.00,25.00){\circle*{2.00}}
\emline{5.00}{45.00}{1}{15.00}{45.00}{2}
\emline{15.00}{45.00}{3}{10.00}{35.00}{4}
\emline{10.00}{35.00}{5}{10.00}{25.00}{6}
\emline{5.00}{45.00}{7}{10.00}{35.00}{8}
\emline{80.00}{10.00}{9}{80.00}{20.00}{10}
\emline{80.00}{20.00}{11}{85.00}{25.00}{12}
\emline{85.00}{25.00}{13}{80.00}{30.00}{14}
\emline{80.00}{30.00}{15}{75.00}{25.00}{16}
\emline{75.00}{25.00}{17}{80.00}{20.00}{18}
\put(35.00,40.00){\circle*{2.00}}
\put(35.00,45.00){\circle*{2.00}}
\put(35.00,50.00){\circle*{2.00}}
\put(35.00,55.00){\circle*{2.00}}
\put(35.00,60.00){\circle*{2.00}}
\put(35.00,65.00){\circle*{2.00}}
\emline{35.00}{65.00}{19}{35.00}{40.00}{20}
\put(45.00,60.00){\circle*{2.00}}
\put(55.00,60.00){\circle*{2.00}}
\put(50.00,55.00){\circle*{2.00}}
\put(50.00,40.00){\circle*{2.00}}
\put(50.00,45.00){\circle*{2.00}}
\put(50.00,50.00){\circle*{2.00}}
\emline{45.00}{60.00}{21}{50.00}{55.00}{22}
\emline{50.00}{55.00}{23}{55.00}{60.00}{24}
\emline{50.00}{55.00}{25}{50.00}{40.00}{26}
\put(68.00,55.00){\circle*{2.00}}
\put(78.00,55.00){\circle*{2.00}}
\put(73.00,50.00){\circle*{2.00}}
\put(65.00,50.00){\circle*{2.00}}
\put(81.00,50.00){\circle*{2.00}}
\put(73.00,43.00){\circle*{2.00}}
\put(98.00,25.00){\circle*{2.00}}
\put(108.00,25.00){\circle*{2.00}}
\put(95.00,20.00){\circle*{2.00}}
\put(111.00,20.00){\circle*{2.00}}
\put(103.00,13.00){\circle*{2.00}}
\emline{65.00}{50.00}{27}{81.00}{50.00}{28}
\emline{68.00}{55.00}{29}{73.00}{50.00}{30}
\emline{73.00}{50.00}{31}{78.00}{55.00}{32}
\emline{73.00}{50.00}{33}{73.00}{43.00}{34}
\put(90.00,55.00){\circle*{2.00}}
\put(100.00,55.00){\circle*{2.00}}
\put(95.00,50.00){\circle*{2.00}}
\put(95.00,40.00){\circle*{2.00}}
\put(95.00,45.00){\circle*{2.00}}
\emline{90.00}{55.00}{35}{95.00}{50.00}{36}
\emline{95.00}{50.00}{37}{100.00}{55.00}{38}
\emline{95.00}{50.00}{39}{95.00}{40.00}{40}
\emline{90.00}{55.00}{41}{100.00}{55.00}{42}
\put(40.00,30.00){\circle*{2.00}}
\put(35.00,25.00){\circle*{2.00}}
\put(35.00,15.00){\circle*{2.00}}
\put(45.00,25.00){\circle*{2.00}}
\put(45.00,15.00){\circle*{2.00}}
\emline{35.00}{15.00}{43}{35.00}{25.00}{44}
\emline{35.00}{25.00}{45}{40.00}{30.00}{46}
\emline{40.00}{30.00}{47}{45.00}{25.00}{48}
\emline{45.00}{25.00}{49}{45.00}{15.00}{50}
\emline{35.00}{25.00}{51}{45.00}{25.00}{52}
\put(55.00,25.00){\circle*{2.00}}
\put(55.00,15.00){\circle*{2.00}}
\put(65.00,25.00){\circle*{2.00}}
\put(65.00,15.00){\circle*{2.00}}
\emline{55.00}{15.00}{53}{55.00}{25.00}{54}
\emline{65.00}{25.00}{55}{65.00}{15.00}{56}
\emline{55.00}{25.00}{57}{65.00}{25.00}{58}
\emline{55.00}{15.00}{59}{65.00}{15.00}{60}
\emline{65.00}{15.00}{61}{55.00}{25.00}{62}
\emline{103.00}{13.00}{63}{111.00}{20.00}{64}
\emline{111.00}{20.00}{65}{108.00}{25.00}{66}
\emline{108.00}{25.00}{67}{98.00}{25.00}{68}
\emline{98.00}{25.00}{69}{95.00}{20.00}{70}
\emline{95.00}{20.00}{71}{103.00}{13.00}{72}
\end{picture}
}
\caption{Graphs of size 4 and 5 that are in $\symm$ but not in $\auto$.}
\label{fig:symgraphs}
\end{figure}

\begin{question}
How much larger is $\symm$ than $\auto$? Are there other connected
separating examples than those listed above?
\end{question}

\section{Closure properties of $\auto$}\label{s:close}

We now recall a few operations on graphs. Given two graphs
$G_1$ and $G_2$, we define a product graph on the vertex set
$\V(G_1)\times \V(G_2)$ in three ways. Two vertices
$(u_1,u_2)$ and $(v_1,v_2)$ are adjacent in the {\em Cartesian
product\/} $G_1\times G_2$ if either $u_1=v_1$ and $\{u_2,v_2\}\in\E(G_2)$
or $u_2=v_2$ and $\{u_1,v_1\}\in\E(G_1)$;
in the {\em lexicographic product\/} $G_1[G_2]$ if either
$\{u_1,v_1\}\in\E(G_1)$ or $u_1=v_1$ and $\{u_2,v_2\}\in\E(G_2)$;
in the {\em categorical product\/} $G_1\cdot G_2$ if
$\{u_1,v_1\}\in\E(G_1)$ and $\{u_2,v_2\}\in\E(G_2)$.

If the vertex sets of $G_1$ and $G_2$ are disjoint, we define
the {\em sum} (or {\em disjoint union}) $G_1+G_2$ to be the graph
with vertex set $\V(G_1)\cup\V(G_2)$ and edge set $\E(G_1)\cup\E(G_2)$.

Using these graph operations, from Example \ref{ex:auto} one can
obtain more examples of graphs in $\symm$.
Note that the class of antipodal graphs itself is closed with
respect to the Cartesian product~\cite{Kot}.

\newpage
\begin{theorem}
\mbox{}
\begin{enumerate}
\item
$\auto$ is closed with respect to the sum and with respect to
the Cartesian, the lexicographic, and the categorical products.
\item
Moreover, $\auto$ is an ideal
with respect to the categorical product, that is, if $G$ is
in $\auto$ and $H$ is arbitrary, then both $G\cdot H$
and $H\cdot G$ are in $\auto$.
\end{enumerate}
\end{theorem}

\begin{proof}
For the sum the claim 1 is obvious. Consider three
auxiliary product notions. Given two graphs
$G_1$ and $G_2$, we define product graphs
$G_1\pr G_2$, $G_1\prr G_2$, and $G_1\prrr G_2$ on the vertex set
$\V(G_1)\times \V(G_2)$ each. Two vertices
$(u_1,u_2)$ and $(v_1,v_2)$ are adjacent in $G_1\pr G_2$ if
$\{u_1,v_1\}\in\E(G_1)$ and $u_2=v_2$; in $G_1\prr G_2$ if
$\{u_1,v_1\}\in\E(G_1)$ and $u_2\ne v_2$; and
in $G_1\prrr G_2$ if $u_1=v_1$ and $\{u_2,v_2\}\in\E(G_2)$.

Given two permutations, $\phi_1$ of $\V(G_1)$ and
$\phi_2$ of $\V(G_2)$, we define a permutation $\psi$
of $\V(G_1)\times \V(G_2)$ by $\psi(u_1,u_2)=(\phi_1(u_1),\phi_2(u_2))$.
If both $\phi_1$ and $\phi_2$ are involutory, so is $\psi$.
If $\phi_1$ and $\phi_2$ are automorphisms of $G_1$ and $G_2$
respectively, then $\psi$ is an automorphism of each $G_1\pri G_2$,
$i=1,2,3$. Finally, it is not hard to see that if both $\phi_1$ and
$\phi_2$ move all edges, so does $\psi$ in each $G_1\pri G_2$, $i=1,2,3$.

Notice now that $\E(G_1\pr G_2)$, $\E(G_1\prr G_2)$, and
$\E(G_1\prrr G_2)$ are pairwise disjoint. Notice also that
$\E(G_1\times G_2)=\E(G_1\pr G_2)\cup\E(G_1\prrr G_2)$ and
$\E(G_1[G_2])=\E(G_1\pr G_2)\cup\E(G_1\prr G_2)\cup\E(G_1\prrr G_2)$.
It follows that if $\phi_1$ and $\phi_2$ are fixed-edge-free
involutory automorphisms of $G_1$ and $G_2$ respectively, then
$\psi$ is a fixed-edge-free involutory automorphism of both
$G_1\times G_2$ and $G_1[G_2]$. Thus, $\auto$ is closed with respect
to the Cartesian and the lexicographic products.

To prove the claim 2, let $G\in\auto$, $\phi$ be a fixed-edge-free
involutory automorphism of $G$, and $H$ be an arbitrary graph.
Define a permutation $\psi$ of $\V(G\cdot H)$ by $\psi(u,v)=(\phi(u),v)$.
It is not hard to see that $\psi$ is a fixed-edge-free involutory
automorphism of $G\cdot H$. Thus, $G\cdot H\in\auto$. The same is true
for $H\cdot G$ because $G\cdot H$ and $H\cdot G$ are isomorphic.
\end{proof}

\begin{example}{\sl $\symm$ is not closed with respect to
the Cartesian, the lexicographic, and the categorical products.}

Denote the first graph in Example \ref{ex:notauto} by $K_3+e$.
The following product graphs are not in $\symm$:
$(K_3+e)\times P_2$, $P_2[K_3+e]$, and $(K_3+e)\cdot(K_3+e)$.
To show this, for each of these graphs we will describe a strategy
allowing \one\/ to destroy an isomorphism between the red and the
blue subgraphs, irrespective of \two's strategy.

$(K_3+e)\times P_2$ has a unique vertex $v$ of the maximum degree 5,
and $v$ is connected to the two vertices $v_1$ and $v_2$ of degree 4 that are connected to each other.
In the first move of a symmetry-breaking strategy, \one\/ chooses
the edge $\{v,v_1\}$. If \two\/ chooses an edge not incident to $v$,
\one\/ creates a star $K_{1,5}$ and wins. If \two\/ chooses an edge
incident to $v$ but not $\{v,v_2\}$, \one\/ chooses $\{v,v_2\}$
and wins creating a triangle $K_3$ in the third move. Assume therefore
that in the second round \two\/ chooses $\{v,v_2\}$.
In the next moves \one\/ creates a star with center at $v$.
If \two\/ tries to create a star with the same center, he loses
because \one\/ can create a $K_{1,3}$ while \two\/ can create at most
a $K_{1,2}$. Assume therefore that in the first four rounds \one\/
creates a $K_{1,4}$ with center at $v$ and \two\/ creates a $K_{1,4}$
with center at $v_2$ (see Figure~\ref{fig:cart}). In the rounds 5--8 \one\/
attaches a new edge to every leaf of the red star. Player \two\/ loses
because he cannot attach any edge to $v$.

\begin{figure}
\centerline{
\setlength{\unitlength}{0.00083333in}
{\renewcommand{\dashlinestretch}{30}
\begin{picture}(3969,2891)(0,-10)
\path(72,71)(72,2546)(3222,2546)
	(3222,71)(72,71)
\path(72,1196)(747,1871)(72,2546)
\path(3222,1196)(3897,1871)(3222,2546)
\dottedline{45}(1647,71)(1647,2546)
\dottedline{45}(72,1196)(3222,1196)
\dashline{75.000}(747,1871)(3897,1871)
\dashline{75.000}(1647,1196)(2322,1871)(1647,2546)
\put(1737,971){\makebox(0,0)[lb]{$v$}}
\put(2412,1646){\makebox(0,0)[lb]{$v_2$}}
\put(1512,2681){\makebox(0,0)[lb]{$v_1$}}
\put(3222,2546){\blacken\ellipse{128}{128}}
\put(3222,2546){\ellipse{128}{128}}
\put(3222,1196){\blacken\ellipse{128}{128}}
\put(3222,1196){\ellipse{128}{128}}
\put(3222,71){\blacken\ellipse{128}{128}}
\put(3222,71){\ellipse{128}{128}}
\put(3897,1871){\blacken\ellipse{128}{128}}
\put(3897,1871){\ellipse{128}{128}}
\put(1647,2546){\blacken\ellipse{128}{128}}
\put(1647,2546){\ellipse{128}{128}}
\put(1647,1196){\blacken\ellipse{128}{128}}
\put(1647,1196){\ellipse{128}{128}}
\put(1647,71){\blacken\ellipse{128}{128}}
\put(1647,71){\ellipse{128}{128}}
\put(2322,1871){\blacken\ellipse{128}{128}}
\put(2322,1871){\ellipse{128}{128}}
\put(72,2546){\blacken\ellipse{128}{128}}
\put(72,2546){\ellipse{128}{128}}
\put(72,1196){\blacken\ellipse{128}{128}}
\put(72,1196){\ellipse{128}{128}}
\put(72,71){\blacken\ellipse{128}{128}}
\put(72,71){\ellipse{128}{128}}
\put(747,1871){\blacken\ellipse{128}{128}}
\put(747,1871){\ellipse{128}{128}}
\end{picture}
}
}
\caption{First four rounds of \one's symmetry-breaking strategy
on $(K_3+e)\times P_2$ (\one's edges dotted, \two's edges dashed,
uncolored edges continuous).}
\label{fig:cart}
\end{figure}
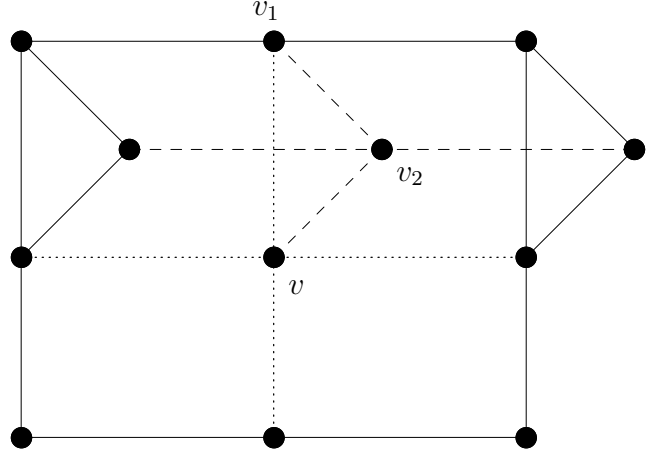

$P_2[K_3+e]$ consists of three copies of $K_3+e$ on the vertex sets
$\{u_1,u_2,u_3,\allowbreak u_4\}$, $\{v_1,v_2,v_3, v_4\}$,
and $\{w_1,w_2,w_3,w_4\}$,
and of 32 edges $\{v_i,u_j\}$ and $\{v_i,w_j\}$ for all $1\le i,j\le 4$
(see Figure~\ref{fig:lexi}). The vertex $v_1$ has the maximum degree 11, $v_2$ and
$v_3$ have degree 10, $v_4$ has degree 9,
and all other vertices have degree at most 7.
In the first move of a symmetry-breaking strategy \one\/ chooses
the edge $\{v_1,v_2\}$. If \two\/ in response does not choose $\{v_1,v_3\}$,
\one\/ does it and wins creating a star with center at $v_1$.
If \two\/ chooses $\{v_1,v_3\}$, in the second move \one\/ chooses
$\{v_1,u_4\}$. If \two\/ then chooses an edge going out of $v_1$,
\one\/ wins creating a $K_{1,6}$. Assume therefore that in the second
move \two\/ chooses an edge $\{v_3,x\}$. If $x=v_2$ or $x=u_4$,
\one\/ chooses $\{u_4,v_2\}$ and wins creating a triangle $K_3$.
Assume therefore that $x$ is another vertex (for example, $x=u_1$
as in Figure~\ref{fig:lexi}). In the third move \one\/ chooses $\{v_2,v_3\}$.
If \two\/ chooses $\{x,v_2\}$, \one\/ chooses $\{u_4,v_3\}$
and wins creating a quadrilateral $C_4$. Otherwise, in the next
moves \one\/ creates a star $K_{1,10}$ with center at $v_2$.
Player \two\/ loses because he can create at most a $K_{1,9}$ with center
at $v_1$ or $v_3$ or at most a $K_{1,7}$ with center at $x$.

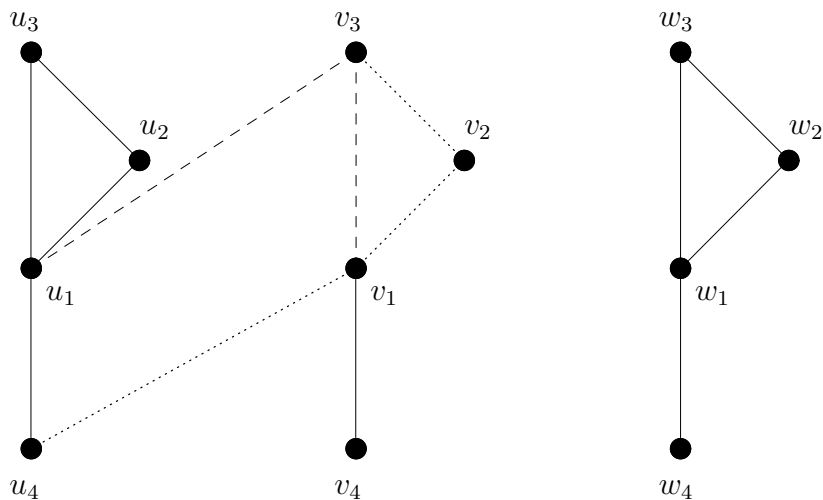
\begin{figure}
\centerline{
\setlength{\unitlength}{0.00083333in}
{\renewcommand{\dashlinestretch}{30}
\begin{picture}(5367,3135)(0,-10)
\path(2160,1440)(2160,315)
\dottedline{45}(135,315)(2160,1440)(2835,2115)(2160,2790)
\dashline{75.000}(135,1440)(2160,2790)(2160,1440)
\path(135,315)(135,2790)(810,2115)(135,1440)
\path(4185,315)(4185,2790)(4860,2115)(4185,1440)
\put(2250,1215){\makebox(0,0)[lb]{$v_1$}}
\put(2025,2925){\makebox(0,0)[lb]{$v_3$}}
\put(2025,0){\makebox(0,0)[lb]{$v_4$}}
\put(225,1215){\makebox(0,0)[lb]{$u_1$}}
\put(0,2925){\makebox(0,0)[lb]{$u_3$}}
\put(0,0){\makebox(0,0)[lb]{$u_4$}}
\put(810,2250){\makebox(0,0)[lb]{$u_2$}}
\put(2835,2250){\makebox(0,0)[lb]{$v_2$}}
\put(4275,1215){\makebox(0,0)[lb]{$w_1$}}
\put(4050,2925){\makebox(0,0)[lb]{$w_3$}}
\put(4050,0){\makebox(0,0)[lb]{$w_4$}}
\put(4860,2250){\makebox(0,0)[lb]{$w_2$}}
\put(2160,2790){\blacken\ellipse{128}{128}}
\put(2160,2790){\ellipse{128}{128}}
\put(2160,1440){\blacken\ellipse{128}{128}}
\put(2160,1440){\ellipse{128}{128}}
\put(2160,315){\blacken\ellipse{128}{128}}
\put(2160,315){\ellipse{128}{128}}
\put(2835,2115){\blacken\ellipse{128}{128}}
\put(2835,2115){\ellipse{128}{128}}
\put(135,2790){\blacken\ellipse{128}{128}}
\put(135,2790){\ellipse{128}{128}}
\put(135,1440){\blacken\ellipse{128}{128}}
\put(135,1440){\ellipse{128}{128}}
\put(135,315){\blacken\ellipse{128}{128}}
\put(135,315){\ellipse{128}{128}}
\put(810,2115){\blacken\ellipse{128}{128}}
\put(810,2115){\ellipse{128}{128}}
\put(4185,2790){\blacken\ellipse{128}{128}}
\put(4185,2790){\ellipse{128}{128}}
\put(4185,1440){\blacken\ellipse{128}{128}}
\put(4185,1440){\ellipse{128}{128}}
\put(4185,315){\blacken\ellipse{128}{128}}
\put(4185,315){\ellipse{128}{128}}
\put(4860,2115){\blacken\ellipse{128}{128}}
\put(4860,2115){\ellipse{128}{128}}
\end{picture}
}
}
\caption{First three moves of \one's symmetry-breaking strategy
on $P_2[K_3+e]$ (\one's edges dotted, \two's edges dashed, uncolored
edges continuous, uncolored edges $\{v_i,u_j\}$, $\{v_i,w_j\}$ not
shown).}
\label{fig:lexi}
\end{figure}

$(K_3+e)\cdot(K_3+e)$ has a unique vertex $v$ of the maximum degree 9,
whereas all other vertices have degree at most 6. A symmetry-breaking
strategy of \one\/ consists in creating a star with center at $v$.
\end{example}

\begin{remark}
$\symm$ is not closed with respect to the sum because, for example,
it does not contain $K_3+P_3$. Nevertheless, if $G_1$ and $G_2$ are
in $\symm$ and both have even size, $G_1+G_2$ is easily seen to be
in $\symm$.
\end{remark}

\section{Complexity of $\auto$}\label{s:np}

Though the graph classes listed in Example \ref{ex:auto} have
efficient membership tests,
%Recognition of membership in $\auto$ of graphs in Example \ref{ex:iso}
%reduces to recognition of graph isomorphism.
in general the existence of an involutory automorphism
without fixed edges is not easy to determine.

\begin{theorem}\label{thm:npcompl}
Deciding membership of a given graph $G$ in the class $\auto$
is NP-complete.
\end{theorem}

\begin{proof}
Consider the related problem \Lubiw\/ whose NP-completeness was
proven in \cite{Lub}.
This is the problem of recognition if a given graph has an involutory
automorphism without fixed {\em vertices}.
We describe a polynomial time reduction $R$ from \Lubiw\/ to $\auto$.

Given a graph $G$, we perform two operations:
\begin{enumerate}
\item[] Step 1.
Split every edge into two adjacent edges by inserting a new vertex, i.e.,
form the subdivision graph $S(G)$ (see \cite[p.\ 80]{Har1}).
\item[] Step 2.
Attach a 3-star by an outer vertex at every non-isolated vertex of $S(G)$
which was in $G$.
\end{enumerate}
As a result we obtain $R(G)$ (see an example in Figure~\ref{fig:red}).
We have to prove that $G$ has an involutory automorphism without fixed
vertices if and only if $R(G)$ has an involutory automorphism without fixed
edges.

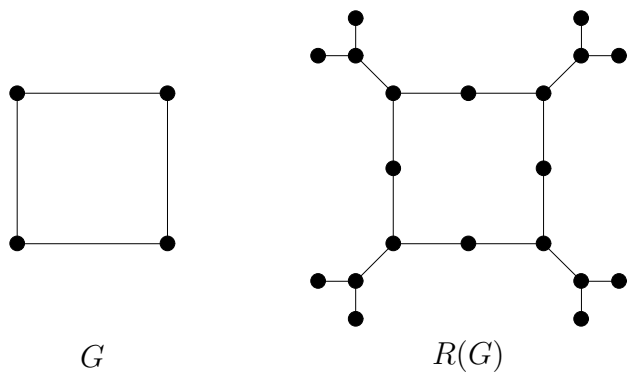
\begin{figure}
\centerline{
\unitlength=1.00mm
%\special{em:linewidth 0.4pt}
\linethickness{0.4pt}
\begin{picture}(86.00,51.00)
\put(5.00,20.00){\circle*{2.00}}
\put(5.00,40.00){\circle*{2.00}}
\put(25.00,40.00){\circle*{2.00}}
\put(25.00,20.00){\circle*{2.00}}
\put(55.00,20.00){\circle*{2.00}}
\put(55.00,40.00){\circle*{2.00}}
\put(55.00,30.00){\circle*{2.00}}
\put(65.00,40.00){\circle*{2.00}}
\put(75.00,40.00){\circle*{2.00}}
\put(75.00,30.00){\circle*{2.00}}
\put(75.00,20.00){\circle*{2.00}}
\put(65.00,20.00){\circle*{2.00}}
\put(50.00,45.00){\circle*{2.00}}
\put(45.00,45.00){\circle*{2.00}}
\put(50.00,50.00){\circle*{2.00}}
\put(80.00,45.00){\circle*{2.00}}
\put(80.00,50.00){\circle*{2.00}}
\put(85.00,45.00){\circle*{2.00}}
\put(50.00,15.00){\circle*{2.00}}
\put(45.00,15.00){\circle*{2.00}}
\put(50.00,10.00){\circle*{2.00}}
\put(80.00,15.00){\circle*{2.00}}
\put(85.00,15.00){\circle*{2.00}}
\put(80.00,10.00){\circle*{2.00}}
\emline{5.00}{20.00}{1}{5.00}{40.00}{2}
\emline{5.00}{40.00}{3}{25.00}{40.00}{4}
\emline{25.00}{40.00}{5}{25.00}{20.00}{6}
\emline{25.00}{20.00}{7}{5.00}{20.00}{8}
\emline{55.00}{20.00}{9}{55.00}{40.00}{10}
\emline{55.00}{40.00}{11}{75.00}{40.00}{12}
\emline{75.00}{40.00}{13}{75.00}{20.00}{14}
\emline{75.00}{20.00}{15}{55.00}{20.00}{16}
\emline{55.00}{20.00}{17}{50.00}{15.00}{18}
\emline{50.00}{15.00}{19}{45.00}{15.00}{20}
\emline{50.00}{15.00}{21}{50.00}{10.00}{22}
\emline{55.00}{40.00}{23}{50.00}{45.00}{24}
\emline{50.00}{45.00}{25}{50.00}{50.00}{26}
\emline{50.00}{45.00}{27}{45.00}{45.00}{28}
\emline{75.00}{40.00}{29}{80.00}{45.00}{30}
\emline{80.00}{45.00}{31}{80.00}{50.00}{32}
\emline{80.00}{45.00}{33}{85.00}{45.00}{34}
\emline{75.00}{20.00}{35}{80.00}{15.00}{36}
\emline{80.00}{15.00}{37}{85.00}{15.00}{38}
\emline{80.00}{15.00}{39}{80.00}{10.00}{40}
\put(65.00,5.00){\makebox(0,0)[cc]{$R(G)$}}
\put(15.00,5.00){\makebox(0,0)[cc]{$G$}}
\end{picture}
}
\caption{An example of the reduction.}
\label{fig:red}
\end{figure}

Every involutory automorphism of $G$ without fixed vertices
determines an involutory automorphism of $R(G)$ that,
thanks to the new vertices, has no fixed edge.
On the other hand, consider an arbitrary automorphism $\psi$ of $R(G)$.
Since $\psi$ maps the set of vertices of degree 1 in $R(G)$ onto itself,
$\psi$ maps every 3-star added in Step 2 into another such 3-star (or itself)
and therefore it maps $\V(G)$ onto itself. Suppose that $u$ and $v$
are two vertices adjacent in $G$ and let $z$ be the vertex inserted
between $u$ and $v$ in Step 1. Then $\psi(z)$ is adjacent in $R(G)$ with
both $\psi(u)$ and $\psi(v)$. As easily seen, $\psi(z)$ can appear
in $R(G)$ only in Step 1 and therefore $\psi(u)$ and $\psi(v)$ are
adjacent in $G$. This proves that $\psi$ induces an automorphism of $G$.
The latter is involutory if so is $\psi$.
Finally, if $\psi$ has no fixed edge, then every 3-star added in Step 2
is mapped to a different such 3-star
and consequently the induced automorphism of $G$ has no fixed vertex.
\end{proof}

Theorem \ref{thm:npcompl} implies that, despite the combinatorial
simplicity of an auto\-mor\-phism-based strategy, realizing this
strategy by \two\/ on $G\in\auto$ requires of him to be at least NP powerful.
The reason is that an automorphism-based strategy subsumes finding an
involutory fixed-edge-free automorphism of any given $G\in\auto$, whereas
this problem is at least as hard as testing membership in $\auto$.

Given the order or the size, there are natural ways of efficiently
generating a graph in $\auto$ with respect to a certain
probability distribution.
Theorem \ref{thm:npcompl} together with such a generating
procedure has two imaginable applications in ``real-life'' situations.

{\em Negative scenario.}
Player \two\/ secretly generates $G\in\auto$ and makes an offer to \one\/ to choose $F$
at his discretion and play the game $\avoid G F$.
If \one\/ accepts, then \two\/, who knows a suitable automorphism of $G$,
follows the automorphism-based strategy and at least does not lose.
\one\/ is not able to observe that $G\in\auto$, unless he can efficiently
solve NP.\footnote{We assume here that \one\/ fails to decide if $G\in\auto$
at least for some $G$. We could claim this failure for {\em most\/} $G$
if $\auto$ would be proven to be complete for the average case \cite{Lev}.}

{\em Positive scenario.}
Player \one\/ insists that before the game an impartial third person,
hidden from \two\/, permutes at random
the vertices of $G$. Then applying the automorphism-based
strategy in the worst case becomes for Player \two\/ as hard
as testing isomorphism of graphs.
More precisely, Player \two\/ faces the following search problem.

\smallskip

\noindent
\PAR\/ ({\sc Permuted Automorphism Reconstruction})\\
{\it Input:} $G$, $H$, and $\beta$, where $G$ and $H$ are isomorphic
graphs in $\auto$, and $\beta$ is a fixed-edge-free involutory
automorphism of $H$.\\
{\it Find:} $\alpha$, a fixed-edge-free involutory automorphism of $G$.

\smallskip

\noindent
We relate this problem to \GI, the {\sc Graph Isomorphism} problem,
that is, given two graphs $G_0$ and $G_1$, to recognize if they are
isomorphic. We use the notion of the Turing reducibility extended
in a natural way over search problems. We say that two problems
are polynomial-time equivalent if they are reducible one to another
by polynomial-time Turing reductions.

%\newpage
\begin{theorem}
The problems \PAR\/ and \GI\/ are polynomial-time equivalent.
\end{theorem}

\begin{proof}
We use the well-known fact that the decision problem \GI\/ is
polynomi\-al-time equivalent with the search problem of finding
an isomorphism between two given graphs \cite[Section 1.2]{KST}.

{\em A reduction from \PAR\/ to \GI.}
We describe a simple algorithm solving \PAR\/ under the
assumption that we are able to construct a graph isomorphism.
Given an input $(G,H,\beta)$ of \PAR, let $\pi$ be an isomorphism
from $G$ to $H$. As easily seen, computing the composition
$\alpha=\pi^{-1}\beta\pi$ gives us a solution of \PAR.

{\em A reduction from \GI\/ to \PAR.}
We will describe a reduction to \PAR\/ from the problem of
constructing an isomorphism between two graphs $G_0$ and $G_1$
of the same size. We assume that both $G_0$ and $G_1$ are
connected and their size is odd. To ensure the odd size, one can
just add an isolated edge to both of the graphs. To ensure the
connectedness, one can replace the graphs with their complements.
If we find an isomorphism between the modified graphs, an isomorphism
between the original graphs is easily reconstructed.

We form the triple $(G,H,\beta)$ by setting $G=G_0+G_1$,
$H=G_0+G_0$, and taking $\beta$ to be the identity map between
the two copies of $G_0$. If $G_0$ and $G_1$ are isomorphic, this is
a legitimate instance of \PAR. By the connectedness of $G_0$ and $G_1$,
if $\alpha\function{\V(G)}{\V(G)}$ is a solution of \PAR\/ on
this instance, it either acts within the connected components
$\V(G_0)$ and $\V(G_1)$ independently or maps
$\V(G_0)$ to $\V(G_1)$ and vice versa. The first possibility
actually cannot happen because the size of $G_0$ and $G_1$
is odd and hence $\alpha$ cannot be at the same time involutory
and fixed-edge-free. Thus $\alpha$ is an isomorphism
between $G_0$ and $G_1$.
\end{proof}

\begin{question}
Is deciding membership in $\symm$ NP-hard?
A priori we can say only that $\symm$ is in PSPACE.  Of course, if the
difference $\symm\setminus\auto$ is decidable in polynomial time, then
NP-completeness of $\symm$ would follow from Theorem \ref{thm:npcompl}.
\end{question}

\section{Game $\avoid{K_n}{P_2}$}\label{s:p_2}

Games on complete graphs are particularly interesting.
Notice first of all that in this case a symmetric strategy is not available.

\begin{theorem}\label{thm:nosymstr}
$K_n\notin\symm$ for $n\ge 4$.
\end{theorem}

\begin{proof}
We describe a strategy of \one\/ that violates the isomorphism
between the red and the blue subgraphs at latest in the
$(n-1)$-th round. In the first two rounds \one\/ chooses two adjacent
edges ensuring that at least one of them is adjacent also to the first
edge chosen by \two. Thus, after the second round the game can be in
one of five positions depicted in Figure~\ref{fig:antisym}.

\begin{figure}
\centerline{
\unitlength=1.00mm
%\special{em:linewidth 0.4pt}
\linethickness{0.4pt}
\begin{picture}(106.00,31.00)
\put(10.00,20.00){\circle*{2.00}}
\put(10.00,30.00){\circle*{2.00}}
\put(20.00,30.00){\circle*{2.00}}
\put(20.00,20.00){\circle*{2.00}}
\put(30.00,20.00){\circle*{2.00}}
\put(40.00,20.00){\circle*{2.00}}
\put(30.00,30.00){\circle*{2.00}}
\put(40.00,30.00){\circle*{2.00}}
\put(50.00,20.00){\circle*{2.00}}
\put(55.00,20.00){\circle*{2.00}}
\put(60.00,20.00){\circle*{2.00}}
\put(50.00,30.00){\circle*{2.00}}
\put(60.00,30.00){\circle*{2.00}}
\put(70.00,20.00){\circle*{2.00}}
\put(80.00,20.00){\circle*{2.00}}
\put(70.00,30.00){\circle*{2.00}}
\put(80.00,30.00){\circle*{2.00}}
\put(90.00,20.00){\circle*{2.00}}
\put(100.00,20.00){\circle*{2.00}}
\put(90.00,30.00){\circle*{2.00}}
\put(100.00,30.00){\circle*{2.00}}
\put(105.00,30.00){\circle*{2.00}}
\emline{20.00}{30.00}{1}{20.00}{20.00}{2}
\emline{20.00}{20.00}{3}{10.00}{30.00}{4}
\emline{30.00}{20.00}{5}{40.00}{20.00}{6}
\emline{40.00}{20.00}{7}{40.00}{30.00}{8}
\emline{55.00}{20.00}{9}{60.00}{30.00}{10}
\emline{60.00}{30.00}{11}{60.00}{20.00}{12}
\emline{70.00}{20.00}{13}{80.00}{30.00}{14}
\emline{80.00}{30.00}{15}{80.00}{20.00}{16}
\emline{100.00}{30.00}{17}{100.00}{20.00}{18}
\emline{100.00}{20.00}{19}{105.00}{30.00}{20}
\put(15.00,10.00){\makebox(0,0)[cb]{1}}
\put(35.00,10.00){\makebox(0,0)[cb]{2}}
\put(55.00,10.00){\makebox(0,0)[cb]{3}}
\put(75.00,10.00){\makebox(0,0)[cb]{4}}
\put(95.00,10.00){\makebox(0,0)[cb]{5}}
\bezier{10}(10.00,20.00)(10.00,25.00)(10.00,30.00)
\bezier{10}(10.00,30.00)(15.00,30.00)(20.00,30.00)
\bezier{10}(30.00,30.00)(35.00,30.00)(40.00,30.00)
\bezier{10}(50.00,30.00)(55.00,30.00)(60.00,30.00)
\bezier{10}(70.00,30.00)(75.00,30.00)(80.00,30.00)
\bezier{10}(90.00,30.00)(95.00,30.00)(100.00,30.00)
\bezier{10}(30.00,20.00)(30.00,25.00)(30.00,30.00)
\bezier{10}(50.00,20.00)(50.00,25.00)(50.00,30.00)
\bezier{10}(70.00,20.00)(70.00,25.00)(70.00,30.00)
\bezier{10}(90.00,20.00)(90.00,25.00)(90.00,30.00)
\end{picture}
}
\caption{First two rounds of \one's symmetry-breaking strategy
(\one's edges dotted, \two's edges continuous).}
\label{fig:antisym}
\end{figure}
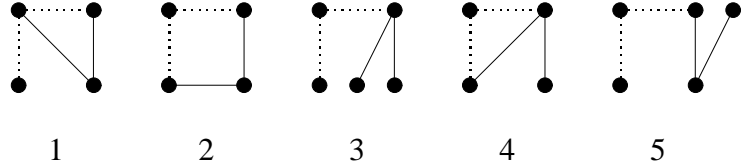

In positions 1 and 2 \one\/ creates a triangle, which is impossible
for \two. In positions 3, 4, and 5 \one\/ creates an $(n-1)$-star, while \two\/
is able to create at most an $(n-2)$-star (in position 5 \one\/ first of all
chooses the uncolored edge connecting two vertices of degree 2).
\end{proof}

Let us define $\win$ to be the class of all graphs $G$ such that,
for all $F$, \two\/ has a non-losing strategy in the game $\avoid GF$.
Clearly, $\win$ contains $\symm$.
It is easy to check that $K_4$ is in $\win$, and therefore
$\symm$ is a proper subclass of $\win$.

It is an interesting question if $K_n\in\win$ for all $n$.
We examine the case of a forbidden subgraph $F=P_2$,
a path of length 2. For all $n>2$, we describe an efficient
winning strategy for \two\/ in $\avoid{K_n}{P_2}$.
Somewhat surprisingly, this strategy, in contrast to
Theorem \ref{thm:nosymstr}, proves to be symmetric in a weaker sense.

More precisely, we say that a strategy of \two\/ is {\em symmetric
in $\avoid GF$\/} if, independently of \one's strategy, the red and
the blue subgraphs are isomorphic after every move of \two\/
in the game. Let us stress the difference with the notion of
a symmetric strategy on $G$ we used so far. While a strategy
symmetric on $G$ guarantees the isomorphism until $G$ is completely
colored (except one edge if $G$ has odd size),
a strategy symmetric in $\avoid GF$ guarantees the isomorphism only
as long as \one\/ does not lose in $\avoid GF$.

\begin{theorem}
Player \two\/ has a symmetric strategy in the game $\avoid{K_n}{P_2}$.
\end{theorem}

\begin{figure}
\centerline{
\unitlength=1.00mm
%\special{em:linewidth 0.4pt}
\linethickness{0.4pt}
\begin{picture}(41.00,26.00)
\put(10.00,10.00){\circle*{2.00}}
\put(10.00,25.00){\circle*{2.00}}
\put(25.00,10.00){\circle*{2.00}}
\put(40.00,10.00){\circle*{2.00}}
\put(40.00,25.00){\circle*{2.00}}
\put(25.00,25.00){\circle*{2.00}}
\emline{10.00}{10.00}{1}{25.00}{25.00}{2}
\emline{25.00}{10.00}{3}{40.00}{25.00}{4}
\emline{10.00}{25.00}{5}{40.00}{10.00}{6}
\bezier{15}(10.00,10.00)(10.00,18.00)(10.00,25.00)
\bezier{15}(25.00,10.00)(25.00,19.00)(25.00,25.00)
\bezier{15}(40.00,10.00)(40.00,17.00)(40.00,25.00)
\end{picture}
}
\caption{A game $(K_6,P_2)$ (\one's edges dotted, \two's edges continuous).}
\label{fig:match}
\end{figure}
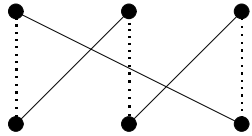

\begin{proof}
Let $A_i$ (resp.\ $B_i$) denote the set of the
edges chosen by \one\/ (resp.\ \two) in the first $i$ rounds.
The strategy
of \two\/ is, as long as $A_i$ is a matching, to choose an edge so that
the subgraph of $K_n$ with edge set $A_i\cup B_i$ is a path.
The only case when this is impossible is that $n$ is even and $i=n/2$.
Then \two\/ chooses the edge that makes $A_i\cup B_i$ a Hamiltonian
cycle (see Figure~\ref{fig:match}).
\end{proof}

\begin{question}
What is the complexity of deciding, given $G$, whether or not \two\/ has
a winning strategy in $\avoid{G}{P_2}$?
\end{question}

It is worth noting that in \cite{Slany99}, PSPACE-completeness of the winner
recognition in the avoidance game with precoloring is proven even for a fixed
forbidden graph $F$, namely for two triangles with a common vertex called the ``bowtie graph''.
Notice also that $\avoid{G}{P_2}$ has an equivalent vertex-coloring
version: the players color vertices of the line graph $L(G)$ and the loser
is the one who creates two adjacent vertices of the same color.

\begin{question}
Does $K_n$ belong to $\win$? In particular, does \two\/ have winning
strategies in $\avoid{K_n}{K_{1,3}}$, $\avoid{K_n}{P_3}$, and
$\avoid{K_n}{K_3}$ for large enough $n$?
\end{question}

\end{document}